\documentstyle[prl,aps,epsf]{revtex}
\draft
\epsfclipon

\begin{document}

\twocolumn[\hsize\textwidth\columnwidth\hsize\csname 
@twocolumnfalse\endcsname
\title{Two-dimensional array of diffusive SNS junctions with
high-transparent interfaces}

\author{T.~I.~Baturina, Z.~D.~Kvon, and A.~E.~Plotnikov}

\address{Institute of Semiconductor Physics, 630090, Novosibirsk, Russia}

\maketitle

\begin{abstract}
We report the first comparative study of the properties of 
two-dimensional arrays and single 
superconducting film--normal wire--superconducting film (SNS)
junctions.
The NS interfaces of our SNS junctions are really high transparent,
for superconducting and normal metal parts are made from the same material
(superconducting polycrystalline PtSi film).
We have found that the two-dimensional arrays reveal some novel features:
(i) the significant narrowing of the 
zero bias anomaly (ZBA) in comparison with single SNS junctions,
(ii) the appearance of subharmonic energy gap structure
(SGS), with up to $n=16$ ($eV=\pm 2\Delta/n$), 
with some numbers being lost,
(iii) the transition from 2D logarithmic weak localization
behavior to metallic one. 

\end{abstract}

]
\bigskip
\narrowtext

Improvements in fabrication technology enable a variety of
nanoscale mesoscopic hybrid structures to be made
and phase-coherent transport to be studied.
In the past few years, the mesoscopic systems, consisting of 
a normal metal (N) or heavily doped semiconductor being 
in contact with
a superconductor (S), have been attracting an increasing
interest mainly because of the richness of involved quantum effects
\cite{1}. 
The key mechanism governing the carrier transport through the NS contact 
is the Andreev reflection.  In this process, an electron-like excitation 
with an energy $\epsilon$ smaller than
the superconducting gap $\Delta$ moving from
the normal metal to the NS interface is retro-reflected as a 
hole-like excitation, while Cooper pair is transmitted into the superconductor. 
This phenomenon is the basis of the proximity effect, 
which generally implies the influence of a 
superconductor on the properties of 
a normal metal being in electrical contact.
Nowadays the comprehensive proximity effect is considered
to be determined by the parameters of the normal part of the junction
and by the properties of the NS interface in common. 
SNS systems may be classified with respect to the
the relation between 
the mean-free path and any other characteristic length in the
system (either diffusive or ballistic regime) 
and by the transparency of the NS interfaces. 
At present, a large variety of NS and SNS junctions
(most of them are diffusive) is fabricated and studied
\cite{2,3,4,5,6,7,8,9,10,11,12,KZD}.
The investigations of these junctions are primarily focused
on the nonlinear behavior
of current-voltage characteristics which exhibit the zero bias anomaly (ZBA),
the subharmonic energy gap structure (SGS), etc. 
Although much work has been done on single SNS junctions, it is an experimental   
challenge to fabricate and carry out comparative measurements on
multiply connected SNS systems. 
In this Letter, we present the results of low-temperature transport 
measurements on two-dimensional
arrays of SNS junctions (2DSNS) and on single SNS junctions 
and perform the comparative analysis of their properties.

The design of our samples is based on the fabrication technique
of SNS junctions with perfect NS interfaces that was recently
proposed and realized by us \cite{KZD}.
The point is the losing superconductivity
in the submicron constrictions made in the ultrathin 
polycrystalline PtSi superconducting film 
by means of electron beam lithography and subsequent plasma etching. 
The reason of that is not completely understood.
Nevertheless, this processing allows us to obtain the SNS junctions,
in which the superconducting and the normal metal parts
are made from the same material.
In that way we escape from unavoidable
uncertainty about parameters of the NS interfaces resulting from
disorder near interface or appearance of oxide barrier that is typical
for intermetallic junctions. 

The original PtSi film with thickness of 6~nm was formed on Si substrate.
The film had critical temperature $T_{\mathrm{c}}=0.56$~K. The resistance
per square was 104~$\Omega$. The carrier density obtained
from Hall measurements was $7\cdot10^{22}$~cm$^{-3}$,
corresponding to the mean-free path $l=1.2$~nm and the diffusion constant 
$D=6$~cm$^2$/s estimated assuming the simple free-electron model.
The initial samples used in the experiments were Hall bridges
with $50~\mu$m in width and $100~\mu$m in length. 

\begin{figure}
\centerline{\epsfxsize3in\epsfbox{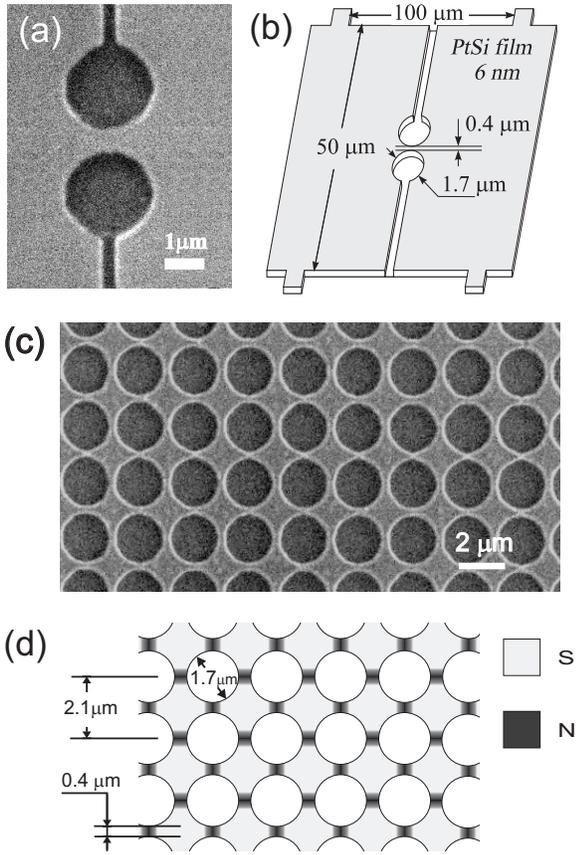}\bigskip}
\caption{(a)~Scanning electron micrograph of the sample with single 
constriction formed by electron beam
lithography and subsequent plasma etching of the $6$~nm PtSi film
grown on Si substrate.
(b)~Schematic view of a junction (not to scale) showing the layout of the 
constriction in the Hall bridge.
(c)~SEM subimage of square lattice of holes made in PtSi film.
(d)~The layout of a two-dimensional array of SNS junctions
showing the dimensions of the sample.
Regions of the normal metal constrictions are dark,
and the superconducting islands are light gray.}
\label{fig1}
\end{figure}

To fabricate 2D array we patterned the square lattice of holes
covering the whole Hall bridge by means of electron lithography
and subsequent plasma etching.
The micrograph of the lattice are shown in 
Fig.~\ref{fig1}c. 
The lattice constant is $2.1~\mu$m
and diameter of holes $1.7~\mu$m,
so the width of the narrowest part
is $0.4~\mu$m.
Thus we obtain the structure which consists of the 
islands of the film, with characteristic dimension being $1.3~\mu$m,
connected by narrow necks. 
As the constrictions
are not superconducting we have a two-dimensional array of
SNS junctions (Fig.~\ref{fig1}d). 
To do a comparative study, 
single SNS junctions were made with the same dimensions
of constrictions, as shown in Fig.~\ref{fig1}a,b.

\begin{figure}
\centerline{\epsfxsize3in\epsfbox{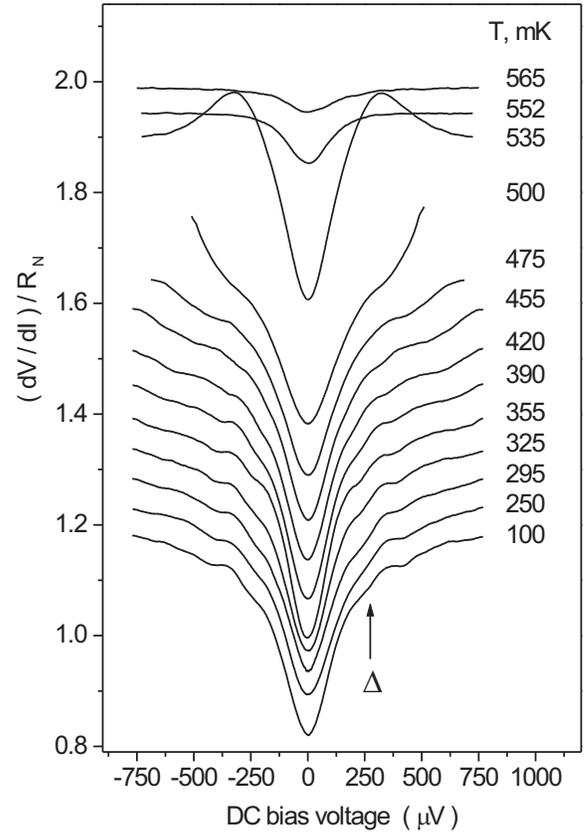}\bigskip}
\caption{Normalized differential resistance $(dV/dI)/R_N$
of a single SNS junction as a function of bias voltage
at different temperatures. (All traces except the lowest trace 
are successively shifted upward by 0.05, for clarity.)
The subharmonic energy gap structure is smeared
and can be seen more clearly on the dependence of 
$d^2V/dI^2$-$V$. The arrow indicates 
$eV=\Delta$ corresponding to the maximum slope.}
\label{fig2}
\end{figure}

The measurements were carried out with the use of a phase sensitive
detection technique at a frequency of 10~Hz that allowed 
us to measure the differential resistance ($R=dV/dI$)
as a function of the dc voltage ($V$).
The ac current was equal to 10~nA.
Figure~\ref{fig2} shows typical dependences of $dV/dI$-$V$ for 
the structures with a single SNS junction. The data exhibit
a behavior very similar to that reported
in the previous work \cite{KZD}. The differential resistance
reveals a minimum at zero bias voltage and reaches $R_N\sim530~\Omega$
at $V_{\mathrm{ZBA}}\sim140~\mu$V ($R_N$ is the difference
between the resistance of the whole structure with constriction
and the resistance of the original film at $T>T_{\mathrm{c}}$).
As one can see from the uppermost curves
obtained at the highest temperatures, the zero bias anomaly survives
even under the condition of fluctuation superconductivity and has
the same voltage scale as that at the lowest temperatures.

\begin{figure}
\centerline{\epsfxsize3in\epsfbox{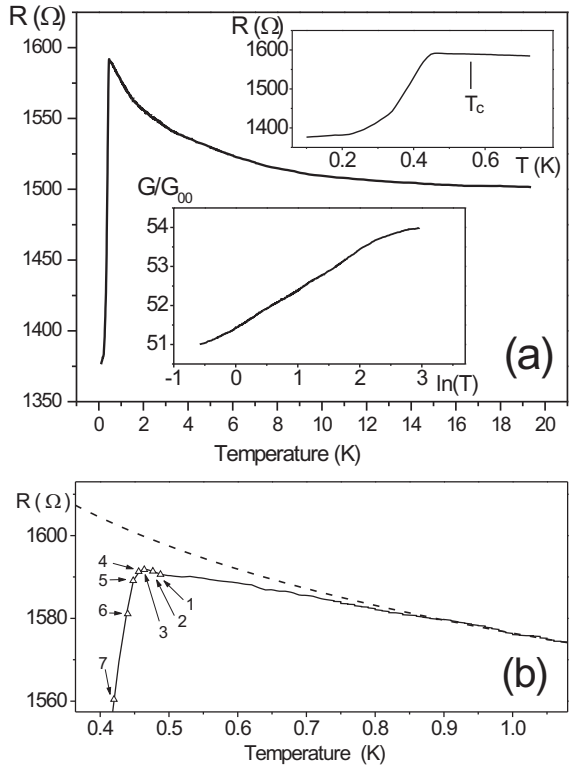}\bigskip}
\caption{(a)~Temperature dependence of the resistance per square for
a two-dimensional array of SNS junctions. The top inset
represents the low-temperature part of the dependence.
The bottom inset shows the conductance in units of 
$G_{00}=e^2/(2\pi^2\hbar)$ vs $\ln(T)$ at $T>T_{\mathrm{c}}$.
(b) The solid curve represents the temperature dependence of the 
resistance nearby
$T_{\mathrm{c}}$. The dashed line is the fit to $1/R(T)\propto\ln(T)$.
The numbered triangles are zero-bias resistance extracted from the 
$dV/dI$-$V$ presented in Fig. 4b (left panel).}
\label{fig3}
\end{figure}

In Fig.~\ref{fig3} we present the temperature dependence of the 
resistance for one of the two-dimensional arrays of SNS junctions.
We have investigated three samples and found the similar behavior. 
At the temperatures higher than the superconducting transition temperature
(this part of the curve is also shown in 
the lower inset to Fig.~\ref{fig3}a)
we observe 
the logarithmic behavior of the conductivity described by
$G(T)=G_{00}\ln(k_BT\tau/\hbar)$, where $G_{00}=e^2/(2\pi^2\hbar)$.
It indicates the effects of the weak localization
and interaction in the diffusion channel in quasi-two-dimensions.
As $T_{\mathrm{c}}$ is approached, the increase of the resistance 
slows down and at a temperature slightly lower than $T_{\mathrm{c}}$
it changes to quick decrease which proceeds down
to the lowest temperature. 
The low-temperature part of the curve is
presented in the top inset to Fig.~\ref{fig3}a.
This behavior can be considered as a transition from 
a 2D logarithmic weak localization to a metallic one due to the
proximity effects. The transition region is depicted in Fig.~\ref{fig3}b.
It should be noted that the resistance of the 
original PtSi film nearby $T_{\mathrm{c}}$ 
is determined by the large contribution
from the superconducting fluctuations. 
Here a deviation from the logarithmic
behavior, originated from the superconducting fluctuations,
becomes apparent at $T\sim0.8$~K. 
As it will be seen from the following results, the rapid 
decrease of the resistance observed at $T\sim0.45$~K, that is less than  
$T_{\mathrm{c}}$ of the unpatterned film, is not due to 
the superconducting transition in the islands. 
The latter occurs at a higher temperature. 
Thus the transition region
is determined by an interplay of several
different contributions: (a)~the weak localization, 
(b)~the superconducting fluctuations, and (c)~the proximity effects.

\begin{figure}
\centerline{\epsfxsize3in\epsfbox{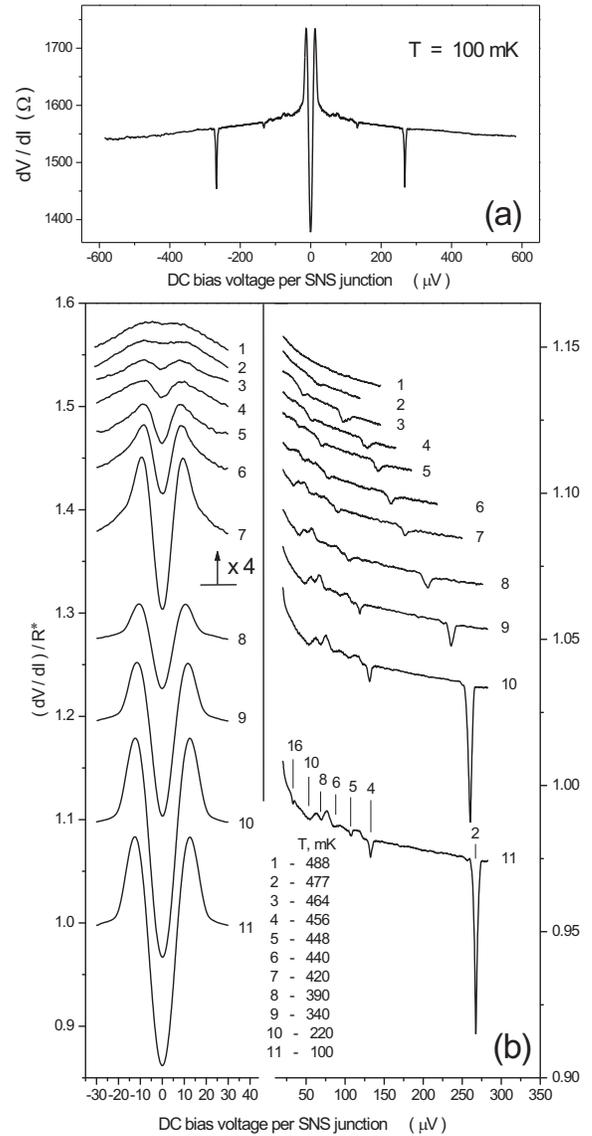}\bigskip}
\caption{(a)~Differential resistance of the sample 
with a two-dimensional array of SNS junctions
as a function of the bias voltage falling at one SNS junction
at $T=100$~mK showing a sharp zero-bias resistance dip followed
by above-normal peak. The differential resistance
possesses the subharmonic energy gap structure that is 
seen as symmetrical minima in a range of voltages from $\sim30~\mu$V
to $\sim300~\mu$V.
(b)~Temperature evolution of the normalized differential resistance
($R^*=1600~\Omega$) vs dc bias voltage divided by the number of SNS
junctions between the measuring probes.
(All traces except the lowest trace 
are shifted up for clarity.)
The left panel shows a low-voltage part of the curves.
(Note the change of height scale in the upper 7 traces.) 
The traces in the right panel are continuations of the ones 
presented in the left panel to higher voltage.
The curves are numbered according to the temperatures listed at the bottom.
}
\label{fig4}
\end{figure}

The differential resistance measurements on 
the two-dimensional array of SNS junctions shown in Fig.~\ref{fig4}
represent the key data of this Letter.
The differential resistance has a minimum at zero bias voltage and
shows a maximum at a finite bias voltage of about $10~\mu$V followed 
by a rapid decrease and eventually a slow decrease at large biases.
In comparison with single SNS junctions where the zero-bias 
resistance dip extends to $\sim140~\mu$V 
for 2DSNS we observe significant narrowing of the ZBA. 
It is less than $10~\mu$V.
At nonzero biases a pronounced and fully
symmetric structure in the differential
resistance is seen. 
It is the so-called subharmonic energy gap
structure (SGS) originated from the multiple Andreev reflections.
In the general case the positions of these features are determined
by the condition $eV=\pm2\Delta/n$, with ($n=1,2,3,\ldots$).
As the right panel in Fig.~\ref{fig4}b shows, the dips corresponding
to $n=2,4,5,6,8,10,16$ manifest in our case.
The presence of SGS in curves 2 and 3 indicates 
clearly 
that in this temperature region 
the islands are already superconducting.
The precise shape of the structure
which we observe for all 2D arrays of SNS
junctions varies from sample to sample, with even $n$
being commonly more pronounced and some $n$ being absent.
Summarizing the observations concerning to the appearance of SGS
in our single and multiply connected SNS systems
one can say that the structure reveals in both systems,
with it being richer in 2DSNS.
It should be noted that 
from the earlier theory \cite{OTBK} describing 
the subgap current transport
in terms of ballistic propagation of quasiparticles
it follows that
there is no chance to observe the SGS in the case of the high-transparent
NS interfaces. 
The SGS on current-voltage characteristics 
of single diffusive SNS junctions was observed in the works
\cite{10,11,12,KZD}
and has only just been explained in a recent paper \cite{difMAR}.
Following the spirit of Nazarov's circuit theory \cite{Nazarov1},
authors of Ref.~\onlinecite{difMAR}
show that unlike the ballistic case in
long diffusive SNS junctions the SGS survives even for
perfect NS interfaces. It occurs owing to coherent 
impurity scattering of the quasiparticles inside the 
proximity region that formally corresponds to renormalized
value of the interface resistance.
 
The issue to be addressed now is the behavior of the differential 
resistance at zero bias. For all samples under study we observe
the dip, with the value of $\Delta R_{\mathrm{SNS}}/R_N$ being 
more than 10\%. 
It is the so-called excess conductance
experimentally observed in all single diffusive SNS junctions
\cite{5,6,7,8,9,10,11,12,KZD}.
ZBA observed in our single SNS junctions is likely
to be the result of interplay of non-local coherent effects,
namely (i)~the superposition of multiple coherent scattering
at the NS interfaces in the presence of disorder
(so-called reflectionless tunneling \cite{RT}) and 
(ii)~the electron-electron interaction in the
normal part.
The latter is one of the important points of recently developed ``circuit theory''
when applied to diffusive superconducting hybrid systems \cite{NazarovStoof}.
Within this approach, which is based 
on the use of the nonequilibrium Green function method,
the electron-electron interaction induces 
weak pair potential in the normal metal. It results not only in the change of 
the resistance, but in a non-trivial distribution of the electrostatic
and chemical potentials in the structure as well, that implies non-local
resistivity in the structure.  
It is essentially a consequence of coherent nature of Andreev reflection. 
The most striking feature is, as the results for 2D array 
of SNS junctions
presented in this Letter show,
that self-averaging is totally absent  
and coherence of the effects governed by Andreev reflection
is maintained over the entire sample. 
Moreover, in comparison with single SNS junction
the manifestation of the effects 
in 2DSNS systems (particularly the SGS) is far more pronounced.
The behavior of the ZBA and SGS in 2D array of SNS junctions
strongly suggests that the development of a 
novel theoretical approach is needed
which would self-consistently take into account 
the distribution of the currents, the potentials, and
the superconducting order parameter.
In this connection a recent work \cite{Lerner} should be noted,
where the authors have extended 
the theoretical approach to disordered systems
based on the nonlinear $\sigma$-model.
An advantage of this approach is, unlike the others where 
the superconducting order parameter was taken into account just 
by the boundary conditions for the normal region, 
that it makes possible
to describe an effect on the superconducting order parameter of
disorder in the normal metal and even inside the superconducting region.
As a consequence it was shown that the size 
of the superconductor influences the proximity effects.
In our case this can be a probable reason for a drastic decrease 
of the effective suppression voltage for the ZBA
when we turn from a single SNS system to multiply
connected SNS junctions. 

In summary, we have performed the first 
comparative study of two-dimensional array 
of diffusive SNS junctions and single SNS junction.
Our experiments show that 
coherent phenomena governed by the Andreev reflection 
are not only maintained over the macroscopic scale but
manifest novel pronounced effects as well. 
To have clear physical understanding of the
phenomena observed in mesoscopic multiply
connected systems further theoretical progress is needed.

We thank R.~Donaton and M.~Baklanov (IMEC) for
providing the PtSi films. 
We would like to acknowledge valuable conversation
with M.~Feigel'man and Yu.~Nazarov.
This work has been supported by the program ``Physics of quantum and
wave processes'' of the Russian Ministry of Science and Technology
and by RFBR Grant No. 00-02-17965.

\end{document}